\documentclass[a4paper,aps,pra,
showpacs,preprintnumbers]{revtex4}
\usepackage{graphics,graphicx,epsfig}
\usepackage[centertags]{amsmath}
\usepackage{amsfonts}

\usepackage{amssymb}
\begin{document}



\title{Broadband squeezed light from phase-locked single-mode sub-Poissonian lasers}

\author{ T.~Golubeva, D.~Ivanov, Yu.~Golubev}
\address{St.~Petersburg State University, V.~A.~Fock Physics Institute, \\
198504 Stary Petershof, St.~Petersburg, Russia}

\date{\today}


\begin{abstract}
We consider sub-Poissonian single-mode laser with external synchronization and analyze its
applicability to the problems of quantum information. Using Heisenberg-Langevin theory we calculate
the quadrature variances of the field emitted by this laser. It is shown that such systems can
demonstrate strong quadrature squeezing. Taking into account that the emitted field is temporally
multi-mode the application of such sources to multichannel quantum teleportation and dense coding
protocols is discussed.
\end{abstract}

\maketitle

\section{Introduction\label{I}}
The  concept  of multi-mode squeezing \cite{MandelWolf} turned out to be very productive for both
quantum optics and quantum information. It is justified by recent advances in quantum holographic
teleportation, ghost imaging \cite{6}, and quantum dense coding \cite{DenseCoding}. These schemes
were based on use of travelling-wave optical parametric amplifiers (OPAs) with spatially multi-mode
structures. Efficiency of these schemes turned out to be essentially enhanced in comparison with
the strongly single-mode models  since many spatial modes ensure the multi-channel parallelism in
the information transfer. In the cited works a temporal spectrum has not been taken into account
because of too wide spectral range of the OPA radiation. Obviously, if systems with similar spatial
structure and reasonably restricted temporal spectrum could be found, then  the efficiency of the
protocols could be made even higher.

From this point of view single-mode lasers seem to be rather perspective. Indeed, on the one hand
there is a possibility to construct from the single-mode laser light a spatially multi-mode
structure. E. g., one could build an array of many lasers which would be able to generate spatial
mode structure not worse (maybe even better) than in the case of the OPA. On the other hand, the
role of the temporal spectrum (the spectral mode width) will be essential too. In realistic lasers
the line width of a mode is similar to spectral capacity of realistic detectors, i. e., the
information encoded in spectral components can be effectively used. At the same time this width is
not small (from units of MHz up to hundreds  of GHz). Thus one can expect that this source of
squeezed light can be more effective than the OPA, since both the spatial and temporal factors
start to play a role.

Below we are going to consider only the temporal aspects of squeezed light from a single-mode
laser. We will show that certain obstacles restricting the direct use of these systems in quantum
information can be overcome. The most important problem is the phase diffusion of a free-running
laser. In quantum information it is extremely important to have a possibility to following the
so-called squeezed quadrature component. However, it is impossible for traditional lasers due to
the phase diffusion, which results in random rotation of the squeezing ellipse. The possible
solution would be to introduce a synchronizing mechanism that would allow for phase diffusion
suppression. However, it is \textit{a priory} not clear whether such synchronization would be
compatible with the quantum features of the sub-Poissonian laser. Detailed investigation of this
problem is the aim of the present paper.

The paper is organized as follows. In Sec.~\ref{II} the Heisenberg-Langevin theory of the
phase-locked sub-Poissonian laser is developed. It is assumed that the laser is synchronized by an
external field in the coherent state.

In order to preserve the quantum features of the laser field we have to limit the power of the
synchronizing field.  Otherwise this field will impose its own Poissonian photon statistics to the
laser mode. It is very clear that too low power is unable to ensure effective phase locking but we
will demonstrate that this restriction is still compatible with an effective laser synchronization.
At the end of the section we calculate spectral densities of the quadrature components of the
intracavity field and express the observed spectral variances via these values. These results will
then be used to evaluate a capacity of the specific information channels.

In the next Secs.~\ref{III} and \ref{IV}, we consider two information schemes, namely the quantum
dense coding and the quantum teleportation with using the sub-Poissonian single mode lasers as the
sources of the multi-mode squeezed light.

\section{Theory of phase-locked single-mode
sub-Poissonian laser\label{II}}
The very first theory of the sub-Poissonian laser was developed in Ref.~\cite{Golubev-jetph}. The
theory was constructed on the basis of the well-known quantum Lamb-Scully approach Ref.~\cite{Lamb}
in terms of the master equation for the field density matrix. Later on this idea was realized
experimentally by Yamamoto et. el. \cite{Yamamoto86} in semiconductor lasers. Now the best results
in  reduction of the shot noises  are reached in low-demensional semi-conductor lasers (in the
so-called VCSELs).

Limited applicability of the sub-Poissonian lasers for the quantum information is due to the phase
diffusion. The random walking of the radiation phase makes it impossible to follow  the squeezed
quadrature components. One of the possibilities to suppress the phase diffusion is to use an
external electromagnetic field in a coherent state for locking the laser phase.  Similar approach
was elaborated, e. g., in Ref.~\cite{WM}, where the problem was considered in terms of the master
equation for the Wigner distribution. Although the main attention there was paid to a role of
feedback, some of the results are important in the context  of the present discussions.

There are two reasons why we want to discuss the theory of the phase-locked laser once more.  First
of all, we would like to have a theory that could be suitable for the semiconductor lasers, because
exactly these systems seem to be the most promising for quantum optics and quantum information. As
is well known the approach based on the master equation and developed in Ref.~\cite{WM} is suitable
only for the gas medium. Here we are going to develop the approach based on the Heisenberg-Langevin
equations. Moreover, the Heisenberg approach turns out to be most convenient for application  of
the theory in the information schemes.

Furthermore, by using the external field in the coherent state for the  phase-locking we have to
remember that the power of the field can not be too high. Otherwise the coherent statistics will be
imposed on the laser field, i. e., the quantum properties of the laser field will be destroyed.
This important question has not been addressed in the previous consideration \cite{WM} but it is
extremely important to find out whether this requirement does not contradict the effective locking
of the laser phase.

\subsection{Physical model and the Heisenberg-Langevin equations }
\begin{figure}[t]
  \centering
  \includegraphics[width=0.4\textwidth]{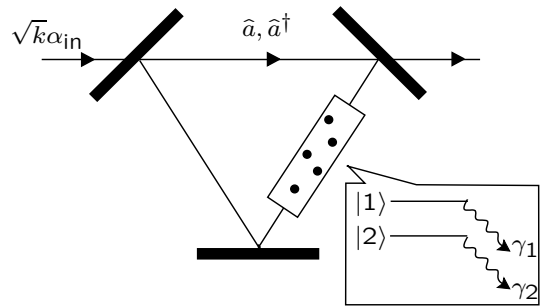}
  \caption{Phase-locked single mode laser}
  \label{fig:model}
\end{figure}

First, we address a physical model of a laser we base our
discussions on (see Fig.~\ref{fig:model}). The principal
components of the model have the properties similar to that of the
standard discussions of the Heisenberg-Langevin laser
theory~\cite{Benkert,Kolobov,Davidovich}. It is assumed that the
laser high-Q cavity supports only a single mode, which is
described by the creation and annihilation operators
$\hat{a}^\dagger$ and $\hat{a}$, obeying the canonic commutation
relation $\left[\hat{a},\hat{a}^\dag\right]\!=\!1$. Furthermore,
one of the mirrors forming the cavity is partially transparent
allowing the laser light to leave the cavity for the
photodetection.

It is assumed that the active laser medium consists of independent two-level atoms with $|1\rangle$ and $|2\rangle$
being the upper and lower lasing levels, respectively.  For the sake of simplicity, the atomic transition between
these levels is assumed to be exactly resonant to the cavity mode. The lifetimes of the atomic states are determined
by the spontaneous decay rates $\gamma_1$ and $\gamma_2$. From the quantum optical point of view the best relation
is $\gamma_1\ll\gamma_2$ \cite{Golubev-jetph}.

Having this model at hand, lasing is described by the following Heisenberg-Langevin equations:
\begin{eqnarray}
&&\dot {\hat a}=-\frac{\kappa}{2}\; \left(\hat a - a_{in} \right)
+ g\hat P+\hat F_a,
\label{A1} \\
&&\dot{\hat P}=-\gamma_\perp{\hat P}+g(\hat N_1-\hat N_2)\hat a+\hat F_p,
\\
&&\dot{\hat N}_1=R-\gamma_1 \hat N_1-g(\hat a^\dag \hat P+\hat a \hat P^\dag)+\hat F_1,
\\
&&\dot{\hat N}_2=-\gamma_2 \hat N_2+g(\hat a^\dag\hat  P+\hat a
\hat P^\dag)+\hat F_2. \label{A4}
\end{eqnarray}
Here $\hat a$ and $\hat a^\dag$ are the mentioned above photon annihilation and creation operators, $\hat{P}$ is the
slowly varying collective atomic polarization on the laser transition, $\hat{N}_1$ and $\hat{N}_2$ are the operators
of the populations of the corresponding atomic states. The coefficients in the equations are: the spectral mode
width $\kappa$, the atom-field coupling constant $g$, the rate of the spontaneous escape $\gamma_1$ and $\gamma_2$,
the rate of the transverse decay $\gamma_\perp$, and the rate of an incoherent pump of the upper laser state $R$.

The inhomogeneous term in the first equation $\kappa a_{in}/2$, where
\begin{eqnarray}
&& a_{in} = \sqrt{n_{in}}\; e^{\displaystyle i \varphi_{in}},
\end{eqnarray}
 ensures a coherent excitation of
  the laser mode by the external classical field with the intracavity amplitude   $a_{in}$.

Apart from the presence of $a_{in}$ the system of equations (\ref{A1})-(\ref{A4}) is exactly the
same as in Ref.~\cite{Benkert,Kolobov,Davidovich}. Note that it is possible to take into account
adiabatically slow phase motion of the synchronizing field assuming $\varphi_{in}=\varphi_{in}(t)$.

The other inhomogeneous terms $\hat F$ in Eqs~(\ref{A1})-(\ref{A4}) represent the noise processes in the laser
system. They appear due to the interaction of the cavity field and the atoms with their own independent Markovian
reservoirs, in both cases being the continuum vacuum modes. These operators possess zero average values and as can
be shown by the direct substitution or via the Einstein relation the only non-zero pair correlation functions read
 \begin{eqnarray}
&&\langle\hat F_a(t)\hat
F_a^\dag(t^\prime)\rangle=\kappa\;\delta(t-t^\prime),
\label{A6}\\
&& \langle\hat F_1(t)\hat F_1(t^\prime)\rangle=\left[\gamma_1
\langle N_1 \rangle+ R(1-p)\right]\;\delta(t-t^\prime),
\\
&& \langle\hat F_2(t)\hat F_2(t^\prime)\rangle=\gamma_2 \langle
N_2 \rangle \;\delta(t-t^\prime),
\\
&& \langle\hat F_p^\dag(t)\hat
F_p(t^\prime)\rangle=\left[(2\gamma_\perp-\gamma_1)\langle N_1
\rangle + R \right] \;\delta(t-t^\prime),
\\
&& \langle\hat F_p(t)\hat
F_p^\dag(t^\prime)\rangle=(2\gamma_\perp-\gamma_2) \langle N_2
\rangle \;\delta(t-t^\prime),
\\
&& \langle\hat F_p(t)\hat F_1(t^\prime)\rangle=\gamma_1 \langle P
\rangle \;\delta(t-t^\prime)
\\
&& \langle\hat F_2(t)\hat F_p(t^\prime)\rangle=\gamma_2 \langle P
\rangle \;\delta(t-t^\prime). \label{A12}
\end{eqnarray}
As is demonstrated in Refs~\cite{Benkert,Kolobov,Davidovich} relying on the model as sketched above one can
introduce the pumping statistics. It can be modelled via distributions of the time instants when excited atoms enter
the cavity. This process is described by the single parameter $ p\le 1$, where $p\!=\!1$ corresponds to the regular
pumping resulting in the sub-Poissonian photon statistics, $p\!=\!0$ to the Poissonian pumping, and $p\!<\!0$ to the
super-Poissonian one.

Aiming to deal with the photodetection one needs not only the mentioned above correlation functions but also their
normally ordered counterparts. These non-zero correlations are obtained in Ref.~\cite{Benkert} and read
 \begin{eqnarray}
&& \langle:\hat F_1(t)\hat
F_1(t^\prime):\rangle=\left[\gamma_1\langle \hat N_1\rangle
-g\;\langle  \hat a^{\dag}\hat P+\hat a
      \hat P^{\dag}\rangle+ R\:(1-p)\right]\;\delta(t-t^\prime),
\label{A13}\\
&& \langle:\hat F_2(t)\hat F_2(t^\prime):\rangle=\left[\gamma_2
\langle N_2 \rangle-g\;\langle  \hat a^{\dag}\hat P+\hat a
      \hat P^{\dag}\rangle\right]
\;\delta(t-t^\prime),
\\
&& \langle:\hat F_1(t)\hat F_2(t^\prime):\rangle=g\;\langle  \hat
a^{\dag}\hat P+\hat a
      \hat P^{\dag}\rangle
\;\delta(t-t^\prime),
\\
&& \langle:\hat F_p^\dag(t)\hat
F_p(t^\prime):\rangle=[(2\gamma_\perp-\gamma_1)\langle N_1 \rangle
+ R \:] \;\delta(t-t^\prime),
\label{eq:pp-corr}\\
&& \langle:\hat F_p(t)\hat F_p(t^\prime):\rangle=2g\;\langle \hat
a\hat P\rangle \;\delta(t-t^\prime),
\\
&& \langle:\hat F_p(t)\hat F_2(t^\prime):\rangle=\gamma_2 \langle
P \rangle \;\delta(t-t^\prime). \label{A18}
\end{eqnarray}
Having specified the model and its theoretical description we proceed with finding the solution applying reasonable
approximations.

In order to solve the system of Eqs~(\ref{A1})-(\ref{A4}), we employ the following commonly used approximations.
First of all we assume that the polarization $\hat P$ and the population $\hat N_2$ of the lower laser level can be
adiabatically eliminated. This is justified given the following requirement
$\gamma_2,\;\gamma_\perp\gg\gamma_1,\;\kappa $ is fulfilled. This relations between the spectral parameters are
typical for the semi-conductor lasers. After this operation the system is essentially simplified, where only two
equations for the population of the upper atomic state $\hat N_1$ and for the field amplitude $\hat a$ survive. The
remaining equations are non-linear differential equations which also cannot be solved analytically too. The next
approximation is based on the assumption that in the stationary regime the system variables only slightly deviate
from the corresponding semiclassical solutions. These approach allows us to linearize equations of motion.

First of all let us apply the adiabatic approximation relative to the fast variables $\hat P$ and
$\hat{N}_2$.
After the adiabatical illumination of these variables the simplified system only for the field amplitude and the
population of the upper laser level is obtained. This system reads
\begin{eqnarray}
&&\dot{\hat N}_1 = - \gamma_1\hat N_1-c\;\hat a^\dag\hat a \hat
N_1+R+\hat \xi_N(t),\qquad c=2g^2/\gamma_\perp,
\label{A19}\\
&&\dot{\hat a}=-\frac{\kappa}{2}\;(\hat
a-a_{in})+\frac{c}{2}\;\hat
 N_1\hat a + \hat{\xi}_a(t), \label{A20}
\end{eqnarray}
where the new noise sources are given by
\begin{eqnarray}
&&\hat{\xi}_a = \hat{F}_a + g/\gamma_\perp\; \hat{F}_p -
g^2/\gamma_2^2\; \hat F_2 \hat{a},\qquad\hat \xi_N=\hat
F_1+c/\gamma_2\;\hat a^\dag\hat a \hat F_2-g/\gamma_\perp\;(\hat
a^\dag \hat F_p+\hat a \hat F_p^\dag).\label{A23}
\end{eqnarray}
We shall find the solutions of the system of Eqs.~(\ref{A19})-(\ref{A20}) assuming that these solutions can be
considered as slow deviations near the semi-classical stationary solutions. Formally this means that quantities
\begin{eqnarray}
 &&\hat a(t)= [\sqrt n  + \delta \hat a(t)]\;e^{i\varphi_{in}},\qquad
  \hat
N_1(t)=N_{1} + \delta \hat N_1(t), \label{22.}
\end{eqnarray}
should satisfy the requirement
\begin{eqnarray}
 && \sqrt n  \gg \delta \hat a(t),\qquad N_{1}\gg \delta \hat
N_1(t). \label{A24}
\end{eqnarray}
It should be  stressed here that the first of the inequalities is valid only if the phase diffusion in the laser is
suppressed.

As for the semi-classical stationary solutions $N_1, \overline a$,
it is not difficult to get that in the only interesting saturation
regime $\gamma_1\ll cn $
\begin{eqnarray}
&&N_1=\frac{R}{\gamma_1}, \qquad  \overline a=\sqrt
n\;e^{\displaystyle i\varphi_{in}}.
\end{eqnarray}
The stationary real amplitude $\sqrt{n}$ satisfies  the following quadratic equation
\begin{eqnarray}
&&\sqrt n\left(\sqrt n-\sqrt
{n_{in}}\right)=\frac{R}{\kappa}\label{26}.
\end{eqnarray}
It is convenient for us to introduce the parameter
\begin{eqnarray}
&&\mu=\sqrt{\frac{n}{n_{in}}},
\end{eqnarray}
which reflects   a role of the external synchronizing factor in a production of the full power of the operation.
Further we will choose that this role is very small, what means that
\begin{eqnarray}
&&\mu\ll1.
\end{eqnarray}
Then in particular Eq.~\ref{26} provides us with the solution $ n\approx R/\kappa$.

\subsection{Linearization of the equations relative to fluctuations}
\begin{figure}[t]
  \centering \includegraphics[width=0.3\textwidth]{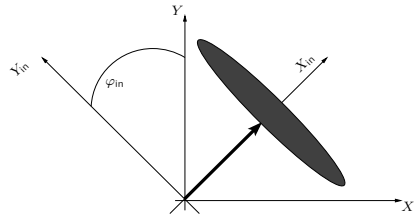}
  \caption{Amplitude and phase quadratures in rotated frame}
  \label{fig:rotation}
\end{figure}

Under requirement (\ref{A24}) one can linearize the adiabatic equations (\ref{A19})-(\ref{A20}) and obtain them in
the following form
\begin{eqnarray}
&&\delta\dot{\hat x}=-\kappa\mu/2\;\delta\hat x+c\sqrt
n/2\;\delta\hat N_1+\hat \xi_x(t),
\label{A33} \\
&&\delta\dot{\hat y}=-\kappa\mu/2\;\delta{\hat y}+\hat\xi_y(t),
\label{A34}\\
&&\dot{\delta\hat N}_1=-\Gamma_1\;\delta\hat N_1-2\kappa(1-\mu)
\sqrt n\;\delta\hat x +\hat\xi_N(t),\qquad\Gamma_1=\gamma_1+cn.
\label{A35}
\end{eqnarray}
Here we have introduced for the further convenience instead of the field amplitude fluctuations $\delta \hat a$ the
Hermitian quadrature operators
\begin{eqnarray}
&&\delta\hat x= \frac{1}{2}(\delta\hat a+ \delta\hat a^\dag),
\qquad\delta\hat y=-\frac{i}{2}(\delta\hat a- \delta\hat a^\dag).
\label{A36}
\end{eqnarray}
Here it needs to stress that these functions are not quite the
quadrature field components inside the cavity. Indeed, according
to (\ref{22.}) we have to take into account an additional
exponential factor coupled with the external field phase. Thus the
quadrature components are expressed by
\begin{eqnarray}
&&\delta\hat x= \frac{1}{2}\left(e^{i\varphi_{in}}\delta\hat a+
e^{-i\varphi_{in}}\delta\hat a^\dag\right), \qquad\delta\hat
y=-\frac{i}{2}\left(e^{i\varphi_{in}}\delta\hat a-
e^{-i\varphi_{in}}\delta\hat a^\dag\right).  \label{32}
\end{eqnarray}
Worth noting that these operators  acquire different physical meaning depending on the phase
$\varphi_{in}$ of the control field. In fact they are the usual amplitude and phase quadratures but
in the coordinate frame rotated by the angle $\varphi_{in}$, see Fig.~\ref{fig:rotation}. One can
see that as $\varphi_{in}=0$ the definitions (\ref{A36}) and (\ref{32}) are perfectly equivalent.

The new noise sources in Eqs.~(\ref{A33})-(\ref{A35}) read
\begin{eqnarray}
&&\hat\xi_x = \frac{1}{2}(e^{-i\varphi_{in}}\hat
F_a+e^{i\varphi_{in}}\hat
F_a^\dag)+\frac{g}{2\gamma_\perp}\;(e^{-i\varphi_{in}}\hat
F_p+e^{i\varphi_{in}}\hat F_p^\dag)-\frac{c\sqrt
n}{2\gamma_2}\;\hat F_2,
\label{A37}\\
&&\hat\xi_y= -\frac{i}{2}(e^{-i\varphi_{in}}\hat
F_a-e^{i\varphi_{in}}\hat
F_a^\dag)-\frac{ig}{2\gamma_\perp}\;(e^{-i\varphi_{in}}\hat
F_p-e^{i\varphi_{in}}\hat F_p^\dag)\\
&&\hat\xi_N = \hat F_1-\frac{g\sqrt
n}{\gamma_\perp}\;(e^{-i\varphi_{in}}\hat
F_p+e^{i\varphi_{in}}\hat F_p^\dag)+\frac{cn}{\gamma_2}\;\hat F_2.
\label{A39}
\end{eqnarray}
To find the partial solution of inhomogeneous linear equations~(\ref{A33})-(\ref{A35}) it is convenient to pass to
Fourier domain. Defining the Fourier transformation for a function $G(t)$  as
\begin{eqnarray}
&& G_\omega=\frac{1}{\sqrt{2\pi}}\int\limits_{-\infty } ^{+\infty
}G(t)\;e^{i\omega t}dt,\qquad
G(t)=\frac{1}{\sqrt{2\pi}}\int\limits_{-\infty } ^{+\infty
}G_\omega\;e^{-i\omega t}d\omega\label{A46}
\end{eqnarray}
one comes to the following equations
\begin{eqnarray}
&&-i\omega\;{\delta\hat N}_{1\omega} = -\Gamma_1\;\delta\hat
N_{1\omega}-2\kappa\;(1-\mu)\sqrt n \;\delta\hat x_{ \omega}
+\hat\xi_{N\omega},
\label{A47}\\
&&-i\omega\;{\delta\hat x_{\omega}} = -\kappa\mu/2\;\delta\hat
x_{\omega}+c/2\;\sqrt n\;\delta\hat N_{1\omega}+\hat \xi_{x\omega}
,
\label{A48}\\
&& - i \omega\;\delta\hat y_{\omega} = -\kappa\mu/2 \;\delta\hat
y_{\omega}+\hat\xi_{y\omega}. \label{A49}
\end{eqnarray}
After some simple algebra the solutions for the field-related fluctuations are found to be
\begin{eqnarray}
&&\delta\hat x_{\omega} = \frac{c\sqrt{n}/2 \;\hat \xi_{N\omega}
  + (\Gamma_1-i\omega) \;\hat \xi_{x\omega}}{(\kappa\mu/2-i\omega)
  (\Gamma_1-i\omega) + cn\kappa(1-\mu) }, \qquad\delta\hat y_{\omega} = \frac{\hat\xi_{y\omega}}{\kappa\mu/2 -
  i\omega} \label{A53},
  \end{eqnarray}
  and the population fluctuation reads
  \begin{eqnarray}
&&\delta\hat N_{1\omega} = \frac{- 2 \kappa(1-\mu) \sqrt{n} \;
\hat\xi_{x\omega} + (\kappa\mu/2 -
  i\omega)\;\hat\xi_{N\omega}}{(\kappa\mu/2 - i\omega)(\Gamma_1 - i\omega) +
  cn\kappa(1-\mu)}.  \label{A54}
\end{eqnarray}
Let us define the spectral densities $(\xi_x^2)_\omega$, $(\xi_y^2)_\omega$, and $(\xi_N^2)_\omega$
of the stochastic sources in Eqs.~(\ref{A33})-(\ref{A35}) as a factors in front of the
delta-functions in the pair correlation functions
\begin{eqnarray}
&&\langle\hat \xi_{x\omega}\;\hat \xi_{x\omega^\prime} \rangle=(
\xi_x^2)_\omega\;\delta(\omega+\omega^\prime),\\
&&\langle\hat \xi_{y\omega}\;\hat \xi_{y\omega^\prime} \rangle=(
\xi_y^2)_\omega\;\delta(\omega+\omega^\prime),\\
&&\langle\hat \xi_{N\omega}\;\hat \xi_{N\omega^\prime} \rangle=(
\xi_N^2)_\omega\;\delta(\omega+\omega^\prime).
\end{eqnarray}
Similarly one can define the spectral densities $(\xi_{y}\xi_{x})_\omega$,
 $(\xi_{x}\xi_{N})_\omega$, $(\xi_{y}\xi_{N})_\omega$ for the cross-correlation

After straightforward algebra one can obtain the following results for the field variances
\begin{eqnarray}
&&(\xi_x^2)_\omega =(\xi_y^2)_\omega = 2i(\xi_{y}\xi_{x})_\omega =-2i(\xi_{x}\xi_{y})_\omega=
\kappa/2\;(1-\mu/2) , \label{A57}\\
&& (\xi_N^2)_\omega=  \kappa(1-\mu)/c \; \Gamma_1 (2-p) ,\label{A55}\\
 &&(\xi_{x}\xi_{N})_\omega =(\xi_{N}\xi_{x})_\omega=-i(\xi_{y}\xi_{N})_\omega =i(\xi_{N}\xi_{y})_\omega= -
\kappa/2\;(1-\mu)\;\sqrt n . \label{A58}
\end{eqnarray}
Now it is possible to derive the spectral intracavity variances in the explicit form
\begin{eqnarray}
&&(\delta x^2)_\omega=\frac{\kappa(1-\mu/2)}{2}
\;\frac{\omega^2+\Gamma_1^2-cn\Gamma_1p/2\;(1-\mu)/(1-\mu/2)}{
(\omega^2-cn\kappa(1-\mu)-\kappa\mu/2\Gamma_1)^2+ \omega^2(\Gamma_1+\kappa\mu/2)^2},
\label{A59}\\
&&(\delta y^2)_\omega=\frac{\kappa(1-\mu/2)}{2}
\;\frac{1}{\omega^2+\kappa^2\mu^2/4} \label{A60}.
\end{eqnarray}
In the saturation regime $\gamma_1\ll cn\to\Gamma_1\approx cn$
\begin{equation}
(\delta x^2)_\omega  = \frac{\kappa}{2}\;
\frac{1-\mu/2-(1-\mu)p/2}{\kappa^2(1-\mu/2)^2+ \omega^2}.
\label{A63}
\end{equation}
We remind that in our theory the synchronizing parameter $\mu$ is much less than one. Nevertheless
we have to survive it in our formulas because further there is compensation in the main order.

\subsection{Spectral field variances outside the cavity}
The single-mode light leaving the cavity for a photodetector is described by the normalized
amplitude $\hat a(t)$ inside  and the other normalized amplitude $\hat A(t)$ outside the cavity.
Normalization is fulfilled such that a value $\langle \hat a^\dag \hat a\rangle$ takes a sense of
the mean photon number inside the cavity and a value $\langle \hat A^\dag \hat A\rangle$ the mean
photon number per sec. through the cross section of the outgoing beam. The boundary condition at
the output mirror can be derive in the form \cite{Gard}
\begin{eqnarray}
&&\hat A(t)=\sqrt\kappa \;\hat a(t)-\hat A_{vac}(t).
\end{eqnarray}
This condition takes into account not only the field leaving the cavity but the vacuum field
reflected by mirror. This ensures that the field amplitude $\hat A(t)$ obeys the canonical
commutation relations
\begin{eqnarray}
&&\left[\hat A(t),\hat
A^\dag(t^\prime)\right]=\delta(t-t^\prime),\qquad\left[\hat
A(t),\hat A(t^\prime)\right]=0
\end{eqnarray}
and it is independent of the processes inside the cavity.

Introducing again the quadrature components for the external field  as real and imaging parts of
the amplitude
\begin{eqnarray}
&&\hat X=\frac{1}{2}\;(\hat A+\hat A^\dag),\qquad\hat
Y=\frac{1}{2i}(\hat A-\hat A^\dag),
\end{eqnarray}
one can couple the spectral variances inside and outside the cavity
\begin{eqnarray}
&&(\delta X^2)_{\omega}=\frac{1}{4}+\kappa\;(:\!\delta
x^2\!\!:)_\omega,\qquad(\delta
Y^2)_{\omega}=\frac{1}{4}+\kappa\;(:\!\delta
y^2\!\!:)_\omega.\label{A62}
\end{eqnarray}
One can see that the external variances are expressed via the normally ordered variances inside the
cavity. The latter can be obtain on the basis the normally ordered variances for the noise sources
\begin{eqnarray}
&&(:\!\xi_x^2\!\!:)_\omega=(:\!\xi_y^2\!\!:)_\omega=\kappa/2(1-\mu),\qquad(:\!\xi_x\xi_N\!\!:)_\omega=
-\kappa(1-\mu) \sqrt n,\qquad(:\!\xi_N^2\!\!:)_\omega= \kappa(1-\mu)/c\;\Gamma_1(2-p)
\end{eqnarray}
and then it is not difficult to get
\begin{eqnarray}
(:\!\delta x^2\!\!:)_\omega &=& \frac{\kappa(1-\mu)}{2}\;
\frac{\omega^2+\Gamma_1^2-cn\Gamma_1(1+p/2)}{(\omega^2-cn\kappa(1-\mu)-\kappa\mu/2\Gamma_1)^2+
\omega^2(\Gamma_1+\kappa\mu/2)^2},  \\
(:\!\delta y^2\!\!:)_\omega &=& \frac{\kappa(1-\mu)}{2}
\;\frac{1}{\omega^2+\kappa^2\mu^2/4}. \label{67}
\end{eqnarray}
In the saturation regime the first variance is simplified to
\begin{eqnarray}
(:\!\delta x^2\!\!:)_\omega &=&- \frac{\kappa(1-\mu)p}{4}\;
\frac{1}{\kappa^2(1-\mu/2)^2+ \omega^2}. \label{A68}
\end{eqnarray}

\subsection{Locking the laser phase }
One of our main goals is to understand how the external synchronization affects the laser radiation properties.
First of all we note that the power of the synchronizing field must not be high, Otherwise it will imposes the
Poissonian photon statistics on the laser radiation, i. e., will destroy the laser squeezing. This is a physical
reason for the made above assumption $\mu\ll1$ that limits the power of the synchronizing field.

However under this restriction we risk to lock the phase not properly. E. g., although the phase is locked but its
fluctuations $\delta\varphi$ could remain high (on the level of one or even more) which renders the laser completely
useless for our aims. Below we will prove that the inequalities $\mu\ll1$ and $\delta\varphi\ll1$ are perfectly
compatible with each other.

\begin{figure}[t]
  \centering \includegraphics[width=0.3\textwidth]{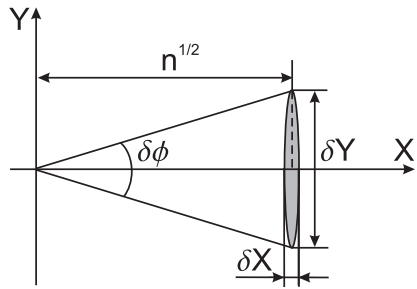}
  \caption{Amplitude and phase variances}
  \label{fig:phase}
\end{figure}

Fig~\ref{fig:phase} will help us to evaluate $\delta\varphi$. In the case $\varphi_{in}=0$ it is
possible to write
\begin{equation}
\langle\delta\varphi^2\rangle=\langle\delta \hat y^2\rangle/{n}
.\label{65}
\end{equation}
The integral y-variance is found from the spectral one by integrating over frequencies
\begin{eqnarray}
&&\langle\delta\hat y^2\rangle=\frac{1}{2\pi }\int(\delta
y^2)_\omega d\omega=\sqrt\frac{n}{4n_{in}}\gg1,
\end{eqnarray}
where the spectral variance $(\delta y^2)_\omega$ is given in the previous subsection (\ref{A60}). Combining this
result with Eq.~(\ref{65}) yields
\begin{equation}
\langle\delta\varphi^2\rangle=\frac{1}{\sqrt{4n_{in}n}}.
\end{equation}
This means that the condition $\langle\delta\varphi^2\rangle\ll1$ is fulfilled always as $n_{in}\gg1/n$. Since we
require that $n\gg1$, the appropriate phase locking is easily realizable without contradiction of any other
requirements. Thus the synchronization by the external field is perfectly effective even for very weak synchronizing
fields, thus the presented analysis is self-consistent.

\section{Quantum dense coding protocol\label{III}}

\subsection{The Duan criterium for continuous variable entanglement }
The Duan criterium \cite{Duan} allows us to predict whether the application of the laser radiation can in principle
be effective for the quantum information purposes. In the case of the multi-mode entanglement, this criterium can be
formulated in the following way: two beams are entangled if there is a frequency region, where the collective
canonic co-ordinates $\delta\hat Q_{1\omega}$ and $\delta \hat Q_{2\omega}$ and the canonic momenta $\delta\hat
P_{1\omega}$ and $\delta \hat P_{2\omega}$ satisfy
\begin{equation}
2\left((\delta Q_{1}+\delta  Q_{2})^2\right)_\omega,\;
2\left((\delta P_{1}-\delta  P_{2})^2\right)_\omega<1 .
\end{equation}
Let us mix two light beams from two independent phase-locked lasers on a symmetrical beamsplitter; then two initial
amplitudes $\hat S_1$ and $\hat S_2$ are modified according to the equalities
\begin{eqnarray}
&&\hat E_{1}=\frac{1}{\sqrt2}(\hat S_{1}+\hat S_{2}),\qquad\hat
E_{2}=\frac{1}{\sqrt2}(\hat S_{1}-\hat S_{2}).\label{81.}
\end{eqnarray}
Introducing the quadrature components  after and before  the beamsplitter according to
\begin{eqnarray}
&&\hat E_{i}=\hat Q_i+i\hat P_i,\qquad\hat S_{i}=\hat X_i+i\hat
Y_i,\qquad i=1,2,\label{81}
\end{eqnarray}
one can obtain for the co-ordinates
\begin{eqnarray}
&&\delta\hat Q_{1,\omega}=\frac{1}{\sqrt2}(\delta\hat
X_{1,\omega}+\delta\hat X_{2,\omega}),\qquad\delta\hat
Q_{2,\omega}=\frac{1}{\sqrt2}(\delta\hat X_{1,\omega}-\delta\hat
X_{2,\omega})
\end{eqnarray}
and for the momenta
\begin{eqnarray}
&&\delta\hat P_{1,\omega}=\frac{1}{\sqrt2}(\delta\hat
Y_{1,\omega}+\delta\hat Y_{2,\omega}),\qquad\delta\hat
P_{2,\omega}=\frac{1}{\sqrt2}(\delta\hat Y_{1,\omega}-\delta\hat
Y_{2,\omega}).
\end{eqnarray}
Due to the assumption that the lasers are statistically independent, we can derive
\begin{eqnarray}
&&\left((\delta Q_{1}+\delta Q_{2})^2\right)_\omega=2(\delta
X_{1}^2)_\omega,\qquad\left((\delta P_{1}-\delta
P_{2})^2\right)_\omega=2(\delta Y_{2}^2)_\omega.
\end{eqnarray}
Now we require that both lasers are identical except the value of the phase $\varphi_{in}$: let us take
$\varphi_{in}=0$ for the first laser  and  $\varphi_{in}=\pi/2$ for the second one. Then taking into account
Eqs.~(\ref{A62})-(\ref{A68}) we obtain the following result
\begin{eqnarray}
&&2\left((\delta Q_{1}+\delta
Q_{2})^2\right)_\omega=2\left((\delta P_{1}-\delta
P_{2})^2\right)_\omega= \frac{\omega^2
+\kappa^2[\mu^2/4+(1-p)(1-\mu)]}{\omega^2+\kappa ^2(1-\mu/2)^2 }.
\end{eqnarray}
In the case of Poissonian lasers (p=0), $2\left((\delta
Q_{1}+\delta Q_{2})^2\right)_\omega=2\left((\delta P_{1}-\delta
P_{2})^2\right)_\omega= 1$ and  we have no any possibility to
think about the entanglement.

However as $p=1$ (sub-Poissonian lasing)
\begin{eqnarray}
&&2\left((\delta Q_{1}+\delta
Q_{2})^2\right)_\omega=2\left((\delta P_{1}-\delta
P_{2})^2\right)_\omega= \frac{\omega^2
+\kappa^2\mu^2/4}{\omega^2+\kappa ^2 }.
\end{eqnarray}
In this regime,  the Duan criterium is well fulfilled for the frequency region $\omega\ll\kappa$.

Thus  we come to the conclusion that the phase-locked sub-Poissonian lasers can serve as an effective source of
entanglement for different protocols of  quantum information. Below we are going to prove it  analizing quantum
dense coding and teleportation protocols.

\subsection{The dense coding setup }
\label{sec:dense-coding}
\begin{figure}[t]
  \centering
  \includegraphics[width=0.6\textwidth]{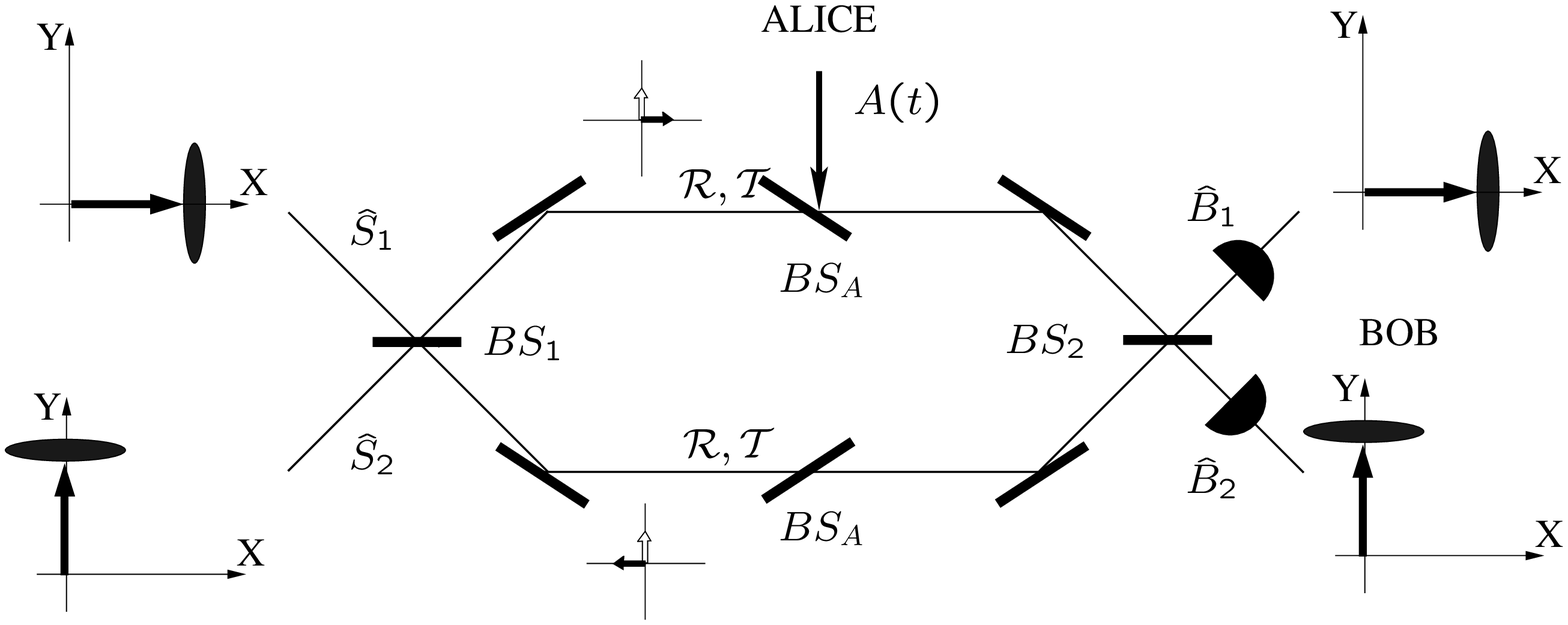}
  \caption{Quantum dense coding protocol}
  \label{fig:coding}
\end{figure}

We will discuss the same setup as in Ref.~\cite{DenseCoding}. The scheme of the protocol
realization is shown in Fig.~\ref{fig:coding}. The setup is based on the Mach-Zehnder
interferometer. All space is divided by optical elements into several domains, therefore it is
convenient to use different letters to name the field amplitudes in different domains. Two sources
of squeezed light emit the beams with amplitudes $\hat S_{1}(t)$ and $\hat S_{2}(t)$. We require
that these sources are perfectly identical but their fields are squeezed in mutually orthogonal
quadratures. Being mixed at the input symmetrical beamsplitter $BS_1$  these fields produce two
beams in an entangled state. We denote the amplitudes of the entangled beams by $\hat E_1(t)$ and
$\hat E_2(t)$ (\ref{81.}). The notation $\hat A(t)$ we reserve for the Alice's signal that is
supposed to be transmitted through the dense coding setup.

The beam $\hat E_1$ in the upper arm of the interferometer interfere at the beamsplitter $BS_A$ with the Alice's
signal $\hat A(t)$ that is intended to be delivered to Bob.  We assume that the Alice's signal comprises the
information-carrying c-number amplitude $A(t)$ (classical signal) and the vacuum noise $\hat A_{vac.1}$:
\begin{eqnarray}
&&\hat A(t)=A(t)+\hat A_{vac.1}(t).
\end{eqnarray}
Thus at  $BS_A$ the field amplitude in the upper arm of the interferometer is transformed according to
\begin{equation}
\hat E_1(t) \to \sqrt{{\cal T}}\hat E_1(t)+\sqrt{{\cal R}}\left(
A(t)+\hat A_{vac.1}(t)\right),
\end{equation}
where ${\cal T}$ and ${\cal R}$ are the transmission and reflection coefficients of $BS_A$, respectively (${\cal
T}+{\cal R}\!\!=\!\!1$).
%

The important property of the Mach-Zehnder interferometer is the ability to restore the quantum state of the input
field at the output. Two independent squeezed beams at $BS_1$ would be the same independent squeezed beams after
$BS_2$ without $BS_A$ in the upper arm. In other words the property to restore quantum states refers to a
symmetrical interferometer. Thus an ancillary beamsplitter identical to $BS_A$ has to be introduced in the lower arm
of the interferometer to preserve the symmetry. Then also the beam $\hat E_2$ will be transformed as
\begin{equation}
\hat E_2(t) \to \sqrt{{\cal T}}\hat E_2(t)+\sqrt{{\cal R}}\hat
A_{vac.2}(t) .
\end{equation}
Finally at Bob's site the beams leaving the interferometer are
\begin{eqnarray}
&&\hat{\tilde{B}}_1(t)=\sqrt{{\cal R}/2}\;\left(A(t)+\hat
A_{vac.1}(t)+\hat A_{vac.2}(t)\right) + \sqrt{{\cal T}}\;\hat S_{1}(t),\label{33}\\
&&\hat{\tilde{B}}_2(t)=\sqrt{{\cal R}/2}\;\left(A(t)+\hat
A_{vac.1}(t)-\hat A_{vac.2}(t)\right) + \sqrt{{\cal T}}\;\hat
S_{2}(t).\label{34}
\end{eqnarray}
One sees that the signal sent by Alice appears in both output beams although it was introduced only in the upper
arm. The next step of the protocol is detecting by Bob of the squeezed quadratures of the output beams in order to
extract the information sent by Alice.

\subsection{Spectral signal-to-noise ratio }
Bob can measure the output fields $\hat B_1(t)$ and $\hat B_2(t)$ with  two independent
photodetectors. Then the corresponding photocurrent operators read
\begin{eqnarray}
&&\hat i_m(t)=\hat B_m^\dag(t)\hat B_m(t),\qquad m=1,2.\label{35}
\end{eqnarray}
In the dense coding scheme one needs, however, to measure the  quadratures, thus a balanced
homodyne detection has to be used. The fluctuation current operators $\delta\hat i_m(t)=\hat
i_m(t)-\langle\hat i_m\rangle$ in this case can be written as
\begin{eqnarray}
&&\delta\hat i_m(t)=\beta_m^\ast\delta\hat
B_m(t)+\beta_m\delta\hat B_m^\dag(t),\label{36}
\end{eqnarray}
where $\beta_m$ is the amplitude of the local oscillator.   us choose the phase of the local
oscillator in  a way such that $\beta_1=\beta_1^\ast\equiv \beta$ and $\beta_2=-\beta_2^\ast\equiv
i\beta$. Then the photocurrent fluctuations read
\begin{eqnarray}
&&\delta\hat i_1(t)=\beta\left(\delta\hat B_1(t)+\delta\hat
B_1^\dag(t)\right),\qquad \delta\hat i_2(t)=i\beta\left(\delta\hat
B_2(t)-\delta\hat B_2^\dag(t)\right).
\end{eqnarray}
This choice ensures that two mutually orthogonal squeezed field quadratures from the different sources
 are selected on the different detectors.

By applying the well-known procedure of calculation, one gets the photocurrent spectrum of each of the
photodetectors in the form
\begin{eqnarray}
&&(\delta i_m^2)_\omega=\beta^2\left[{\cal R} +  {\cal T}4(\delta
X_1^2)_\omega+{\cal R}\;\sigma^A_\omega\right].
\end{eqnarray}
Here we again assume that both lasers forming the EPR light are completely identical but they are synchronized to
have different phases zero and $\pi/2$ respectively. This means that $(\delta X_1^2)_\omega=(\delta Y_2^2)_\omega$
and $(\delta Y_1^2)_\omega=(\delta X_2^2)_\omega$. Defining the signal-to-noise ratio as
\begin{eqnarray}
&&{\bf SN\!R}_{\;\omega}=\frac{{\cal R}\;\sigma^A_\omega}{{\cal
R}+{\cal T}\;4(\delta X_1^2)_\omega},
\end{eqnarray}
it is not difficult to derive for this quantity the following result in the saturation laser regime
\begin{eqnarray}
&&{\bf SN\!R}_\omega=\frac{{\cal
R}\sigma^A_\omega\;[\omega^2+(1-\mu/2)^2\kappa ^2
]}{\omega^2+(1-\mu/2)^2\kappa ^2 -{\cal T}p\;(1-\mu)\;\kappa
^2}\;.\label{88}
\end{eqnarray}
Performing this calculation we assumed that the Fourier component $A_\omega$ of the Alice amplitude obeys the
Gaussian statistics and the probability density for this value reads
\begin{equation}
W(A_\omega)=\frac{1}{\pi\sigma^A_\omega}\;
\exp\left(-\frac{|A_\omega|^2}{\sigma^A_\omega}\right).
\end{equation}
Further under the calculation of the Shannon information we will choose the explicit form for the variance
$\sigma^A_\omega$ as
\begin{equation}
\sigma^A_\omega=\frac{P}{\sqrt{\pi\Delta\omega_A^2/2}}\;
\exp\left(-\frac{\omega^2}{\Delta\omega_A^2/2}\right),\label{87}
\end{equation}
where $P$ is the integral Alice stream dense (the mean photon numbers per $\sec$) and $\Delta\omega_A$ is the
spectral width of Alice's signal.

If the laser operates in the Poissonian regime ($p=0)$, then the signal is observed on the level of
the shot noise. We regard this value of the signal-to-noise ratio as the minimal one. It reads
\begin{eqnarray}
&&{\bf SN\!R}_{\;\omega}^{\;min}={\cal
R}\;\sigma^A_\omega.\label{46}
\end{eqnarray}
  In the sub-Poissonian regime  as $p=1$ the spectral variance becomes  frequency dependent
\begin{eqnarray}
&&{\bf SN\!R}_\omega=\frac{\omega^2+\kappa ^2 }{\omega^2+({\cal
R}+\mu^2/4)\kappa ^2 }\;{\cal R}\sigma^A_\omega.\label{88}
\end{eqnarray}
and achieve its maximum as $\Delta\omega_A\ll\kappa$.

The factor in front of the minimal signal-to-noise ratio contains three parameters: ${\cal R}$ and
${\cal T}$ (${\cal R}+{\cal T}=1$) determine the  input possibilities of Alice and $\mu\ll1$
restricts the power of the synchronizing field in the laser. It is readily seen that this factor
can be made very large as ${\cal R}\ll1$ and the laser operates in the appropriate regime.

\subsection{Shannon mutual information}
To calculate the information capacity due to the dense coding protocol we apply the same theoretical consideration
as in Refs~\cite{DenseCoding}. As discussed in these references the appropriate quantitative
characteristic in this case is the so-called Shannon mutual information ({\it SMI}) stream density. Assuming a
Gaussian information channel, one can express the {\it SMI} stream density for this channel as
\begin{eqnarray}
&& { I}^{Sh} = \int\limits_{-\infty}^{+\infty}\ln \left( 1 + {\bf
SN\!R}_{\;\omega}\right)d\omega.
\end{eqnarray}
Substituting in here Eq.~(\ref{88}) and choosing the spectral variance for  Alice's signal as the  Gaussian
distribution (\ref{87}) we can calculate the {\it SMI}.

In the case of the Poissonian laser (p=0), {\it SMI} stream density reads
\begin{equation} { I}^{Sh} =\int\limits_{-\infty}^{+\infty} \ln \Bigl[ 1 +
\frac{{\cal
    R}
P}{\sqrt{\pi
    \Delta\omega_A^2/2}}\;e^{\displaystyle-\frac{\omega^2}{\Delta\omega_A^2/2}}\Bigr]d\omega.
\end{equation}
For a weak Alice signal as ${\cal R} P\ll\Delta\omega_A$ this integral can be explicitly evaluated
to give
\begin{eqnarray}
&&{ I}^{Sh}={\cal R} P,
\end{eqnarray}
 i. e., the {\it SMI}  stream density is exactly equal to the Alice signal stream density in the
information channel.

\begin{figure}[t]
  \centering
  \includegraphics[width=0.4\textwidth]{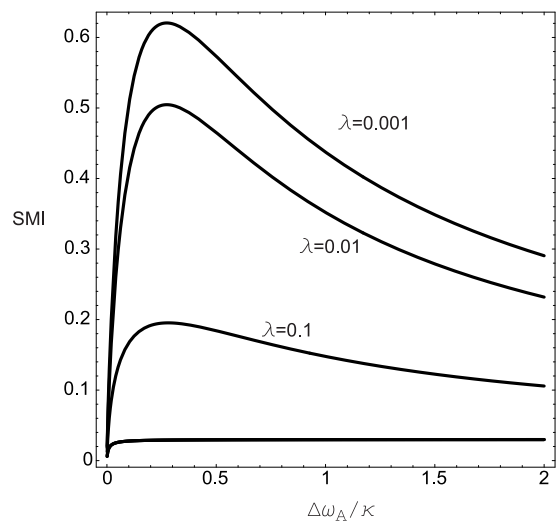}
  \caption{Frequency dependence of the mutual information stream under
  the parameters $\sqrt{\cal R}=0.1,\;{\cal P}=2\pi P/\kappa=3 $}
  \label{fig:stream}
\end{figure}

Being interested in quantum regimes we therefore choose the sub-Poissonian laser regime $p=1$. Then a combined
expression for the \textit{SMI} can be written in the following form
\begin{equation} { I}^{Sh}=\int\limits_{-\infty}^{+\infty} \ln \Bigl[ 1 +
\frac{\omega^2+\kappa ^2}{\omega^2+\kappa ^2({\cal
    R}+\mu^2/4)}\; \frac{{\cal
    R} P}{\sqrt{\pi
    \Delta\omega_A^2/2}}\;e^{\displaystyle-\frac{\omega^2}{\Delta\omega_A^2/2}}\Bigr]
    d\omega.\label{smi_las_dopo}
\end{equation}
To perform numerical integration in Eq~(\ref{smi_las_dopo}) it is convenient to introduce the dimensionless
parameters
\begin{equation}
2\pi I^S/\kappa={\cal I}^S\qquad\omega/\kappa\to\omega,\qquad
2\pi\Delta\omega_A/\kappa=d_A,\qquad 2\pi P/\kappa={\cal P}.
\end{equation}
The results of the integration are shown in Fig.~\ref{fig:stream}. All the curves represent the dependence of the
\textit{SMI} on the scaled bandwidth $d_A$. The lower curve corresponds to the Poissonian regime  of the laser
$(p=0)$ while all other curves to the sub-Poissonian regime $(p=1)$ with different $\lambda=\mu^2/4$ belonging to
the interval from 0.1 to 0.001. The uppermost curve corresponds to $\lambda=0.001$. Note that further decreasing of
this parameter does not shift the curve. The reason for this is the fact that $\lambda$-parameter appears in the
equations only in the combination $\lambda+{\cal R}$, where we have chosen ${\cal R}=0.01$. Thus the information
transfer can be further improved decreasing the reflection of the $BS_A$.

From the figure, one can see the essential advantage of using the sub-Poissonian laser regime. In this case the SMI
can achieve the value of 0.6, while for information transfer with classical light one has only 0.04. This difference
depends on the value $P/\kappa$ that has been chosen to be equal to 3. If $P/\kappa$ is increased, then the
difference becomes less pronounced and practically disappears at $P/\kappa=3000$.

\section{Quantum teleportation protocol\label{IV}}
\subsection{General scheme}
\begin{figure}[t]
  \centering
  \includegraphics[width=0.5\textwidth]{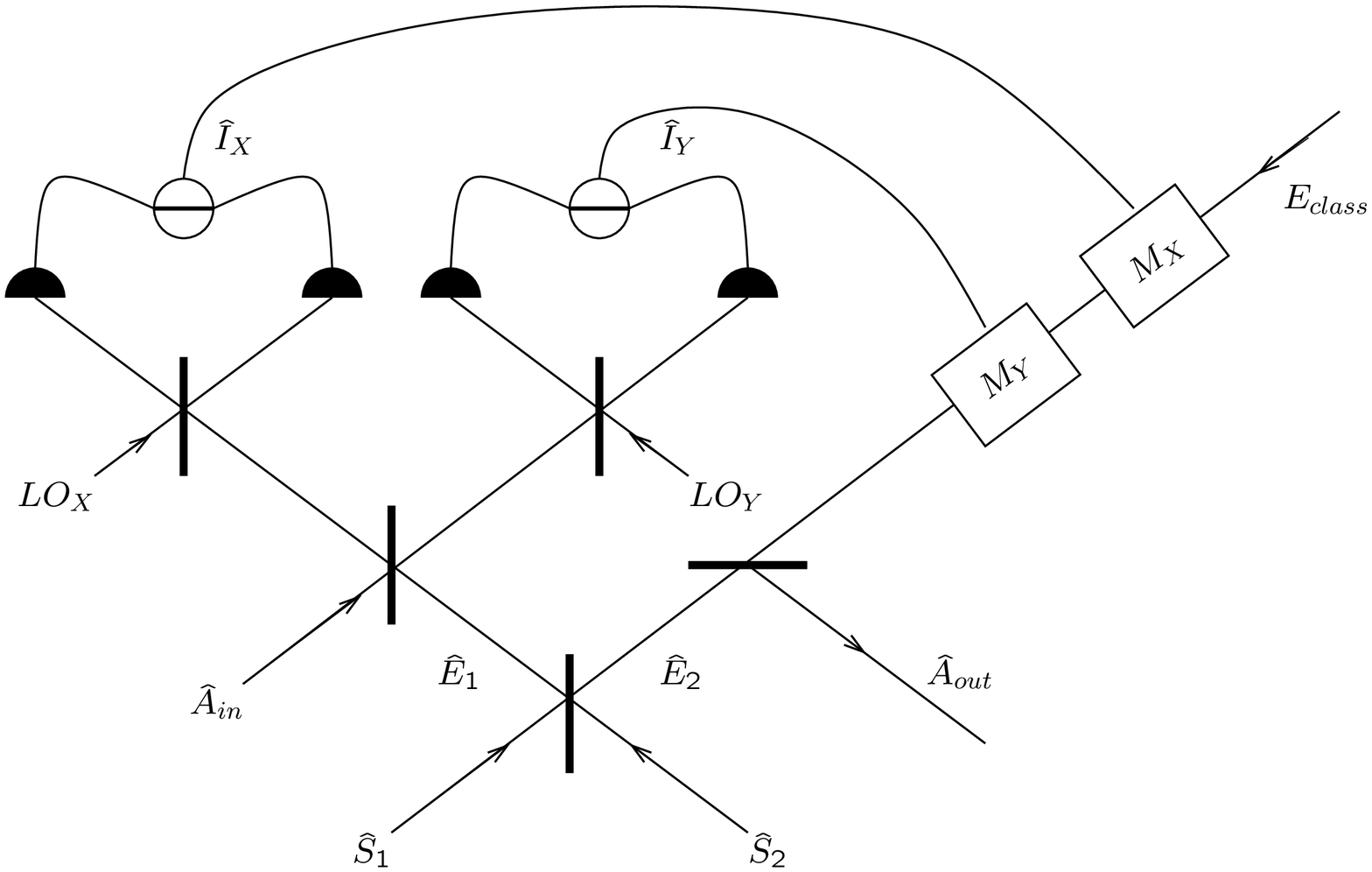}
  \caption{Quantum teleportation protocol.}
  \label{fig:teleport}
\end{figure}

We shall discuss the optical scheme for the teleportation similar to that proposed and realized in
\cite{6}. This scheme is shown in Fig~\ref{fig:teleport}. As for the dense coding the amplitudes
$\hat{S}_1(t)$ and $\hat{S}_2(t)$ presenting the independent laser beams are converted on the
symmetrical beam splitter to $\hat{E}_1(t)$ and $\hat{E}_2(t)$. The last amplitudes describe the
entangled two beam light. One of the beams is mixed again on the symmetrical beamsplitter with the
Alice signal  $\hat A_{in}(t)$:
\begin{eqnarray}
&&\hat B_x(t)=\frac{1}{\sqrt2}\left(\hat A_{in}(t)+\hat
E_1(t)\right),\qquad\hat B_y(t)=\frac{1}{\sqrt2}\left(-\hat
A_{in}(t)+\hat E_1(t)\right).
\end{eqnarray}
Then with the help of appropriately chosen local oscillators $LO_X$ and $LO_Y$ the amplitude quadrature of the field
$\hat{B}_x(t)$ and the phase quadrature of $\hat{B}_y(t)$ are detected. This in particular means that the
photocurrents
\begin{eqnarray}
&&\hat i_x(t)=\beta\left(\hat B_x^\dag(t)+\hat
B_x(t)\right),\qquad\hat i_y(t)=i\beta\left(\hat B_y^\dag(t)-\hat
B_y(t)\right)
\end{eqnarray}
are registered. These photocurrents are sent from Alice to Bob via two classical communication lines. Bob uses these
photocurrents for reconstruction of the field $\hat{A}_{out}(t)$ via two modulators $M_x$ and $M_y$ which modulate
the corresponding quadrature components of an incoming plane coherent light wave with the mean amplitude $E_0$. At
the output of the modulators the amplitude reads
\begin{equation}
\hat E(t) = E_0 + \xi\left(\hat i_x(t)-i\;\hat i_y(t)\right) ,
\end{equation}
where $\xi$ describes the efficiency of the modulation. Mixing this field with the other part of
the EPR-pair on the well reflecting beamsplitter (${\cal R} \!\approx \!1, \, {\cal T}\!\ll\!
1,\;{\cal R}+{\cal T}=1$) Bob obtains his copy of Alice's signal
\begin{eqnarray}
&&\hat A_{out}(t)=\hat A_{in}(t)+\hat F(t),\qquad\hat
F(t)=\sqrt2\left(\delta\hat X_1(t)-i\;\delta\hat Y_2(t)\right)
\end{eqnarray}
given the following equalities hold
\begin{equation}
\xi\beta \sqrt{2{\cal T}}=1,\qquad \sqrt {\cal T}/2\; E_0+ \langle X_1\rangle-i \langle
Y_2\rangle=0.
\end{equation}
Here we introduce the quadrature component $\hat X$ and $\hat Y$ according to
\begin{eqnarray}
&&\hat S_m=\hat X_m+i\hat Y_m,\qquad m=1,2
\end{eqnarray}
and their fluctuations
\begin{eqnarray}
&&\hat X_m=\langle\hat X_m\rangle+\delta\hat X_m,\qquad \hat Y_m=\langle\hat Y_m\rangle+\delta\hat
Y_m.
\end{eqnarray}

Thus the output field completely reproduces the input one if the noise contribution $\hat{F}(t)$ is
negligibly small. Using the quadrature squeezed states of the laser radiation  one reduces $\delta
\hat{X}_1$ and $\delta \hat{Y}_2$, thereby increasing the fidelity of the teleportation.

In the Fourier domain the operator at the output of the teleportation scheme reads
\begin{eqnarray}
&&\hat A_{out,\omega}=\hat A_{in,\omega}+\hat F_\omega,\qquad F_\omega=\sqrt2\;(\delta\hat
X_{1,\omega}-i\;\delta\hat Y_{2,\omega}) .
\end{eqnarray}

\subsection{Spectral fidelity of the teleportation protocol}
Addressing fidelity of a teleportation scheme we deal with optical fields propagating in free
space, which can be presented as a set of oscillators parameterized by their frequencies $\omega$.
We denote by $\hat\rho^{in}_\omega$ and $\hat\rho^{out}_\omega$ the density operators  of the
$\omega$-oscillator at the input and at the output of our teleportation scheme, respectively. Let
the spectral fidelity of this scheme is defined as
\begin{eqnarray}
&&    {\cal F}_\omega= {\rm Tr} \left(\hat\rho_\omega^{in}\;
\hat\rho_\omega^{out} \right)/{\rm Tr}
\left(\hat\rho_\omega^{in}\right)^2.
\end{eqnarray}
For the pure state this expression is converted into well known
$|\langle\psi_{in}|\psi_{out}\rangle|^2$.

Using the formalism of the Wigner function this definition can be cast into the form containing ordinary functions
\begin{eqnarray}
&&{\rm Tr} \left(\hat\rho_\omega^{in}\; \hat\rho_\omega^{out}
\right)=\pi\int d^2\alpha_{\omega}\;
W^{in}_\omega(\alpha_{in,\omega})W^{out}_\omega(\alpha_{out,\omega}),\qquad{\rm
Tr} \left(\hat\rho_\omega^{in} \right)^2=\pi\int
d^2\alpha_{\omega}
\left(W^{in}_\omega(\alpha_{in,\omega})\right)^2. \label{98}
\end{eqnarray}
Let the Wigner distributions can be factorized in the form
\begin{eqnarray}
&&
W^{in,out}_\omega(\alpha_{in,out\omega})=W^{in,out}_\omega(X_{in,out\omega})\;W^{in,out}_\omega(Y_{in,out\omega}),
\end{eqnarray}
where
\begin{eqnarray}
&&\qquad\alpha_\omega=X_\omega+iY_\omega.
\end{eqnarray}
Let the quadrature component fluctuations obey the Gaussian distribution; then the integrals
(\ref{98}) can be calculated  and the spectral fidelity is expressed via the in,out-variances in
the explicit form
\begin{eqnarray}
&& {\cal F}_\omega = \left[\frac{1 }{1+(\delta
X_{1}^2)_\omega/(\delta
X_{in}^2)_\omega}\right]^{\frac{1}{2}}\left[\frac{1}{1+(\delta
Y_{2}^2)_\omega/(\delta Y_{in}^2)_\omega}\right]^{\frac{1}{2}}
\end{eqnarray}

If one teleports light in a coherent state then $(\delta X_{in}^2)_\omega=(\delta Y_{in}^2)_\omega=1/4$ and the
fidelity reads
\begin{eqnarray}
&& {\cal F}_\omega = \left[\frac{1 }{1+4(\delta
X_{1}^2)_\omega}\right]^{\frac{1}{2}}\left[\frac{1}{1+4(\delta
Y_{2}^2)_\omega}\right]^{\frac{1}{2}}
\end{eqnarray}
Putting then $(\delta X_{1}^2)_\omega=(\delta Y_{2}^2)_\omega$, and using the result for the quadrature spectral
variance of the sub-Poissonian laser one ends up with the following result for the spectral fidelity
\begin{eqnarray}
&& {\cal F}_\omega = \frac{1}{2}\;\frac{\omega^2+\kappa ^2 }{\omega^2+\kappa ^2(1-p/2)},\qquad
F_{\omega=0} = \frac{1 }{2-p}.
\end{eqnarray}
For the laser in Poissonian regime $(p=0)$ the expected result $ F_\omega=1/2$ is revealed. This is just the
classical limit for the teleportation fidelity however, for the sub-Poissonian lasing $(p=1)$ the fidelity appears
to
\begin{eqnarray}
&& F_\omega = \frac{1}{2}\;\frac{\omega^2+\kappa ^2
}{\omega^2+\kappa ^2/2}.
\end{eqnarray}
One can see that at zero frequency $F_{\omega=0}=1$, which quantitatively indicates the improved
quality of the teleportation with non-classical light and proves the feasibility of the proposed
sources.

\section{Summary\label{V}}

Aiming to develop an effective source of  entangled light we consider in this paper quantum
properties of radiation emitted by the sub-Poissonian laser with external synchronization. It has
been shown that these systems can operate in essentially non-classical regimes and hence can be
used in temporally multi-mode quantum information schemes. In particular, we have shown that weak
external synchronization can greatly reduce laser phase diffusion without disturbing non-classical
photon statistics. This allows for generation of quadrature-squeezed light. Moreover, varying the
phase of the synchronizing field one can squeeze different quadratures. Thus we believe that the
laser source discussed in this paper can serve as an effective and flexible tool for quantum
information.

To prove the feasibility of the considered systems the operation of two well-known quantum
information protocols based on these systems has been studied. First, the quantum dense coding has
been addressed, where the Shannon mutual information  of the channel based on the phase-locked
sub-Poissonian laser has been calculated. It has been shown that the regimes for the laser exist,
where the Shannon information considerably exceeds that for the channel based on a coherent beam.
This regime is realized for regular pumping (sub-Poissonian generation) and weak synchronizing
external field, i.e., in the case of best squeezing.

The proposed systems have been also tested in a quantum teleportation scheme. To quantify the efficiency of the
systems in the teleportation protocol the spectral fidelity has been considered. It has been shown that operating in
the discussed above essentially quantum regimes both system demonstrate the zero-frequency fidelity of 2/3 whereas
the teleportation with classical field yields 1/2. Furthermore, the results for the information protocols have been
confirmed on a more general basis of applying the Duan entanglement criterium.

\section{ACKNOWLEDGEMENT\label{VI}}

This work was performed within the French-Russian cooperation program "Lasers and Advanced Optical
Information Technologies"with financial support from the following organizations: INTAS (Grant No.
7904), YS-INTAS (Grant No. 6078), RFBR (Grant No. 05-02-19646), and Ministry  of education and
science of RF (Grant No. RNP 2.1.1.362). We would like to thank M. Kolobov and C. Fabre for
productive discussions.

\newpage


 \begin{figure}[t]
  \centering
 \includegraphics[width=0.2\textwidth]{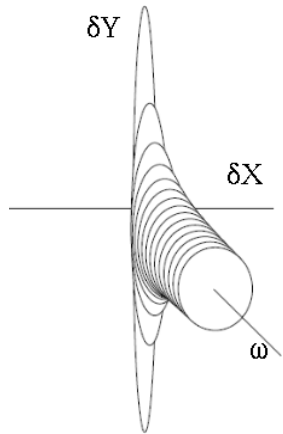}
  \caption{The uncertainty ellipse versus the frequency of the mode}
  \label{fig:circ-ellips}
\end{figure}

\begin{references}
%
\bibitem{MandelWolf} L.~Mandel, E.~Wolf,{\it Optical Coherence and Quantum
Optics}, Cambridge University Press, New York (1995)
%
\bibitem{6}
In collection "Quantum Imaging" ed. by M. Kolobov,  Springer (2007)
%
\bibitem{DenseCoding} T.~Yu.~Golubeva, Yu.~M.~Golubev,
I.~V.~Sokolov, and M.~I.~Kolobov, J. Mod. Opt., {\bf 53}, 699–711 (2006)
%
\bibitem{Golubev-jetph}
Y.~M.~Golubev and I.~V.~Sokolov, Sov.~Phys.~JETP \textbf{60}, 234 (1984)
\bibitem{Lamb}
Scully M. O., Lamb W. E., Phys. Rev., A159, 208 (1967)
%
\bibitem{Yamamoto86} Y.~Yamamoto, S.~Machida, and O.~Nilsson, Phys.~Rev.~A
\textbf{34}, 4025 (1986)
%
\bibitem{WM} H.M.~Wiseman and G.J.~Milburn,
  Phys. Rev. A, \textbf{49}, 1350 (1994)
%
  %
\bibitem{Benkert} C.~Benkert, M.~O.~Scully, J.~Bergou, L.~Davidovich,
M.~Hillery, and M.~Orszag, Phys.~Rev.~A \textbf{41}, 2756 (1990)
%
\bibitem{Kolobov} M.~I.~Kolobov, L.~Davidovich, E.~Giacobino, and C.~Fabre,
Phys.~Rev.~A \textbf{47}, 1431 (1993)
%
\bibitem{Gard}M. J. Collett and C. W. Gardiner
Phys. Rev. A 30, 1386 (1984)
%
\bibitem{Davidovich} L.~Davidovich, Rev.~Mod.~Phys. \textbf{68}, 127 (1996)
%
\bibitem{Duan}Lu-Ming~Duan, G.~Giedke, J.J.~Cirac, and P.~Zoller,
Phys.~Rev.~Lett.\textbf{84}, 2722-2725 (2000).
%



  \end{references}
\end{document}